\documentclass[10pt,a4paper]{article}

\usepackage[utf8]{inputenc}
\usepackage[english]{babel}

\begin{document}
\large

\newpage
\begin{center}
\LARGE {\bf Spin Polarization Type Dependence of the Neutrino 
Mass and Nature}
\end{center}
\large
\vspace{0.1cm}
\begin{center}
{\bf B. S. Yuldashev and R. S. Sharafiddinov}
\end{center}
\vspace{0.1cm}
\begin{center}
{\bf Institute of Nuclear Physics, Uzbekistan Academy of Sciences,
\\Tashkent, 100214 Ulugbek, Uzbekistan}
\end{center}
\vspace{0.1cm}

\begin{verse}
\noindent
{\bf Abstract.} The interaction of longitudinal and transversal polarized 
neutrinos (antineutrinos) with the field of a nucleus is investigated at 
the account of their rest mass, charge, magnetic, anapole and electric dipole 
moments. The compound structure of cross sections in these processes has the 
sharply expressed features and generalities for any lepton as well as for a 
massive Majorana neutrino. A new influence of truly neutral neutrino masses 
on the elastic scattering by nuclei of an electric charge has been discovered 
which testifies in favor of the existence of fundamental differences both in the 
nature and in the masses of longitudinal and transversal neutrinos of Majorana.
\end{verse}
\vspace{0.4cm}

\noindent
One of highly characteristic features of all types of neutrinos is their 
rest mass. At the same time, the nature of inertial mass itself thus far 
remains not finally studied. Usually it is accepted that no any connection 
between the mass of a particle and its fundamental structure. However, 
in the framework of the hypothesis of field mass based on the classical 
theory of an extensive electron [1], a particle all the mass has an 
electromagnetic behavior.

On the other hand, it is known that massless neutrinos are strictly longitudinally 
polarized. In many articles, the properties of longitudinal Dirac neutrinos were
investigated with the account of their rest mass. At the availability of a non-zero 
mass, the neutrino can have either longitudinal or transversal polarization. In this 
appears a sharp dependence between the mass of the neutrino and its spin nature [2].

The purpose of a given work is to express the idea more clearly and to generalize 
their to the case of truly neutral neutrinos studying the interaction of massive 
neutrinos of a different nature with the field of emission in the spin polarization 
type dependence. At first, the processes of elastic scattering of longitudinal 
polarized Dirac $(\nu=\nu_{D})$ and Majorana $(\nu=\nu_{M})$ neutrinos 
(antineutrinos) on a spinless nucleus have been considered taking into account 
the leptonic current charge $F_{1\nu},$ magnetic $F_{2\nu},$ anapole $G_{1\nu}$ 
and electric dipole $G_{2\nu}$ form factors. Next, we will reanalyze their for 
the transversal case of fermion polarization. Some implications implied from 
these discussions have been listed which allow one to follow the behavior of 
neutrinos of both types in the nuclear Coulomb field.

The amplitude of elastic scattering of arbitrary polarized neutrinos
on the nucleus electric charge may be written in the form
\[M^{em}_{fi}=
\frac{4\pi\alpha}{q^{2}}\overline{u}(p',s')
[\gamma_{\mu}F_{1\nu}(q^{2})-
i\sigma_{\mu\lambda}q_{\lambda}F_{2\nu}(q^{2})+\]
\begin{equation}
+\gamma_{5}\gamma_{\mu}G_{1\nu}(q^{2})-
i\gamma_{5}\sigma_{\mu\lambda}q_{\lambda}G_{2\nu}(q^{2})]
u(p,s)J_{\mu}^{\gamma}(q),
\label{1}
\end{equation}
where $\nu=\nu_{D}=\nu_{e L,R}$ or $\nu_{M}=\nu_{1 L,R},$ $q=p-p',$ $p(s)$ 
and $p'(s')$ imply the four-momentum (helicities) of the neutrino before and 
after the interaction, $J_{\mu}^{\gamma}$ denotes the current of a nucleus.

According to these data, the studied processes at the account of longitudinal 
polarization of fermions are described by the cross section
\[\frac{d\sigma_{em}^{V_{\nu},A_{\nu}}(\theta,s,s')}{d\Omega}=
\frac{1}{2}\sigma^{\nu}_{o}(1-\eta^{2}_{\nu})^{-1}
\{(1+ss')\times\]
\[\times[F_{1\nu}
\pm 2s\sqrt{1-\eta_{\nu}^{2}}G_{1\nu}]F_{1\nu}
ctg^{2}\frac{\theta}{2}+\]
\[+\eta^{2}_{\nu}(1-ss')[F_{1\nu}^{2}+
4m_{\nu}^{2}(1-\eta^{-2}_{\nu})^{2}F_{2\nu}^{2}]-\]
\[-8sE_{\nu}^{2}(1-ss')(1-\eta_{\nu}^{2})^{3/2}
F_{2\nu}G_{2\nu}+\]
\[+(1-\eta^{2}_{\nu})[(1+ss')G_{1\nu}^{2}
ctg^{2}\frac{\theta}{2}+\]
\begin{equation}
+4E_{\nu}^{2}(1-ss')G_{2\nu}^{2}]\}
Z^{2}F_{c}^{2}(q^{2})tg^{2}\frac{\theta}{2}.
\label{2}
\end{equation}
Here the upper (lower) sign corresponds to the neutrino (antineutrino),
$V_{\nu}$ and $A_{\nu}$ imply the presence both of vector and of axial-vector 
components of leptonic current, $F_{c}$ is the charge ($F_{c}(0)=1$) form factor 
of a nucleus, $Z$ is the proton number. We have also used the sizes
\[\sigma_{o}^{\nu}=
\frac{\alpha^{2}cos^{2}\frac{\theta}{2}}{4E_{\nu}^{2}
(1-\eta^{2}_{\nu})sin^{4}\frac{\theta}{2}}, \, \, \, \,
\eta_{\nu}=\frac{m_{\nu}}{E_{\nu}},\]
where $\theta$ is the scattering angle, $E_{\nu}$ and $m_{\nu}$ are 
the neutrino mass and energy.

If we take into account that the terms $(1+ss')$ and $(1-ss')$ characterize
the scattering with $(s'=-s)$ and without $(s'=s)$ flip of the spin of incoming 
left $(s=-1)$- and right $(s=+1)$-polarized particles, the value of (\ref{2}) 
one can replace to
\begin{equation}
d\sigma_{em}^{V_{\nu},A_{\nu}}(\theta,s)=
d\sigma_{em}^{V_{\nu},A_{\nu}}(\theta,s'=s)+
d\sigma_{em}^{V_{\nu},A_{\nu}}(\theta,s'=-s),
\label{3}
\end{equation}
in which the first cross section corresponds to a longitudinal neutrino 
scattering with its helicity conservation. The second term is responsible for 
the interconversion of longitudinal polarized neutrinos of different components.

Basing on (\ref{1}), one can define the general structure of the discussed 
process cross sections for the transversal case of neutrino polarization:
\[\frac{d\sigma_{em}^{V_{\nu},A_{\nu}}(\theta,\varphi,s,s')}{d\Omega}=
\frac{1}{2}\sigma^{\nu}_{o}(1-\eta^{2}_{\nu})^{-1}
\{((1+ss')\alpha_{T}cos^{2}\frac{\varphi}{2}+\]
\[+(1-ss')\alpha^{*}_{T}sin^{2}\frac{\varphi}{2})F_{1\nu}^{2}+
\eta^{2}_{\nu}((1+ss')\gamma_{T}sin^{2}\frac{\varphi}{2}-\]
\[-(1-ss')\gamma^{*}_{T}cos^{2}\frac{\varphi}{2})
[F_{1\nu}^{2}+4m_{\nu}^{2}(1-\eta^{-2}_{\nu})^{2}F_{2\nu}^{2}]
tg^{2}\frac{\theta}{2}\pm\]
\[\pm 2s\eta_{\nu}\sqrt{1-\eta_{\nu}^{2}}
((1+ss')sin^{2}\frac{\varphi}{2}-
(1-ss')cos^{2}\frac{\varphi}{2})
\gamma^{*}_{T}F_{1\nu}G_{1\nu}tg\frac{\theta}{2}+\]
\[+(1-\eta^{2}_{\nu})((1+ss')\alpha^{*}_{T}sin^{2}\frac{\varphi}{2}+
(1-ss')\alpha_{T}cos^{2}\frac{\varphi}{2})G_{1\nu}^{2}+\]
\[+4E_{\nu}^{2}(1-\eta^{2}_{\nu})
((1+ss')\gamma^{*}_{T}cos^{2}\frac{\varphi}{2}-\]
\begin{equation}
-(1-ss')\gamma_{T}sin^{2}\frac{\varphi}{2})
G_{2\nu}^{2}tg^{2}\frac{\theta}{2}\}Z^{2}F_{c}^{2}(q^{2}).
\label{4}
\end{equation}
Here it has been accepted that
\begin{equation}
\alpha_{T}=1-2(1-4sin^{2}\frac{\varphi}{2})
sin^{2}\frac{\varphi}{2},
\label{5}
\end{equation}
\begin{equation}
\alpha^{*}_{T}=
1+2(1-4sin^{2}\frac{\varphi}{2})cos^{2}\frac{\varphi}{2},
\label{6}
\end{equation}
\begin{equation}
\gamma_{T}=1+2cos^{2}\frac{\varphi}{2},\,\,\,\, \gamma^{*}_{T}=
1-2cos^{2}\frac{\varphi}{2},
\label{7}
\end{equation}
and $\varphi$ is the azimuthal angle.

As well as in (\ref{2}), the presence in (\ref{4}) of the multipliers $(1+ss')$
and $(1-ss')$ gives the right to present the latter in the form
\begin{equation}
d\sigma_{em}^{V_{\nu},A_{\nu}}(\theta,\varphi,s)=
d\sigma_{em}^{V_{\nu},A_{\nu}}(\theta,\varphi,s'=s)+
+d\sigma_{em}^{V_{\nu},A_{\nu}}(\theta,\varphi,s'=-s),
\label{8}
\end{equation}
where the first term responds to a transversal neutrino scattering without flip 
of its spin. The second cross section characterizes the change of transversal 
polarized neutrino helicities.

In the case of massive Dirac neutrinos, the cross sections (\ref{3}) and 
(\ref{8}) at the low energies $(E_{\nu_{D}}\rightarrow m_{\nu_{D}})$ are 
not different. Unlike this, a coincidence of cross sections for the longitudinal 
and transversal polarized neutrinos of an arbitrary energy takes place if an 
elastic scattering arises at the expense of either vector or axial-vector 
components of leptonic current.

Such an order, however, may exist only in the case where a fermion mass 
is nohow connected with its spin polarization. Therefore, if it turns out 
that the availability of mass in the neutrino transforms its longitudinal 
polarization into the transversal one, and vice versa, this will indicate to 
the existence of fundamental differences not only in the masses [2], but also 
in the nature of Dirac neutrinos of the most diverse types of polarizations.

Exactly the same one can as a completeness include in the discussion the
massive Majorana neutrinos. They have no neither an electric charge nor a 
magnetic moment $(F_{i\nu_{M}}=0),$ and the interaction of their axial-vector 
currents $G_{i\nu_{M}}$ with the field of emission is stronger [3] than of 
Dirac neutrinos. 

Our analysis shows that the cross sections describing the studied processes 
with longitudinal and transversal polarized Majorana fermions of any energy 
have the same size.

It is clear, however, that the same neutrino must possess either longitudinal or 
transversal polarization. Therefore, at the availability of a non-zero mass, the 
longitudinal neutrino can be converted into the transversal one, and vice versa. 
Such a connection between the truly neutral neutrinos in their spin polarization 
type dependence may serve as an indication to the existence of fundamental 
differences both in the nature and in the masses of longitudinal and transversal 
neutrinos of Majorana.
\vspace{0.4cm}

\noindent
{\bf References}
\begin{enumerate}
\item
E. Fermi, Rend. Lincei, 31, 184, 306 (1922); Phys. Zeit. 23, 340 (1922).
\item
R.S. Sharafiddinov, in Proc. Int. Conf. on Nuclear Physics, St-Petersburg, 
June 14-17, 2000 (St-Petersburg, 2000), p. 121.
\item
R.S. Sharafiddinov, Spacetime Subst. 5, 32 (2004); hep-ph/0306145.
\end{enumerate}

\end{document}